 \documentclass[preprint2]{aastex}

\slugcomment{Not to appear in Nonlearned J., 45.}

\shorttitle{Properties of solar polar coronal plumes}
\shortauthors{Raouafi et al.}

\begin{document}


\title{Properties of solar polar coronal plumes constrained by
the Ultraviolet Coronagraph Spectrometer data}


\author{N.-E. Raouafi\altaffilmark{1}, J. W. Harvey\altaffilmark{1} and S. K. Solanki\altaffilmark{2}}
\affil{$^1$National Solar Observatory (NSO), P.O. Box 26732, Tucson, AZ 85726, USA}
\affil{$^2$Max-Planck-Institut f\"ur Sonnensystemforschung,
             Max-Planck-Stra$\ss$e 2, 37191 Katlenburg-Lindau, Germany}




\begin{abstract} We investigate the plasma dynamics (outflow speed and turbulence) inside polar
plumes. We compare line profiles (mainly of \ion{O}{6}) observed by the UVCS instrument on SOHO at
the minimum of solar cycle 22-23 with model calculations. We consider Maxwellian velocity
distributions with different widths in plume and inter-plume regions. Electron densities are
assumed to be enhanced in plumes and to approach inter-plume values with increasing height.
Different combinations of the outflow and turbulence velocity in the plume regions are considered.
We compute line profiles and total intensities of the \ion{H}{1}~Ly$\alpha$ and the \ion{O}{6}
doublets. The observed profile shapes and intensities are reproduced best by a small solar wind
speed at low altitudes in plumes that increases with height to reach ambient inter-plume values
above roughly 3-4 $R_\sun$ combined with a similar variation of the width of the velocity
distribution of the scattering atoms/ions. We also find that plumes very close to the pole give
narrow profiles at heights above 2.5~$R_\sun$, which are not observed. This suggests a tendency for
plumes to be located away from the pole. We find that the inclusion of plumes in the model
computations provides an improved correspondence with the observations and confirms previous
results showing that published UVCS observations in polar coronal holes can be roughly reproduced
without the need for large temperature anisotropy. The latitude distributions of plumes and
magnetic flux distributions are studied by analyzing data from different instruments on SOHO and
with SOLIS. \end{abstract}





\section{Introduction}

This paper is a forward modeling investigation of the plasma dynamics of polar plumes. These
so-called ``polar rays'' or polar plumes (e.g., van de Hulst 1950; Saito 1958; Harvey 1965;
Koutchmy, 1977; DeForest et al. 1997) trace open magnetic field lines to different heights
depending on observation type. In Extreme UltraViolet (EUV) emission line spectroheliograms, plumes
appear as short spikes near the polar limb (Bohlin et al. 1975a; Ahmad \& Withbroe 1977; Widing \&
Feldman 1992; Walker et al. 1993) with unchanged shape and sudden brightness changes over typical
lifetimes ranging from several hours to days (see DeForest et al. 1997; Del~Zanna et al. 1997). EUV
observations can be used to characterize the bases of polar plumes and also their plasma
characteristics (densities, temperatures, flows, abundances, etc.) and also other processes such as
heating, wave propagation, etc. (e.g. Bohlin et al. 1975b; Walker et al. 1988, 1993). Soft X-rays
show mainly the hot plasma at the foot-points of the plumes (e.g., Ahmad \& Webb, 1978) that are
also seen as weak radio sources (Gopalswamy et al., 1992). DeForest et al. (1997, 2001a) have shown
that most white light polar rays can be associated directly with the EUV polar plumes and their
foot-points on the solar surface.

Above the solar limb, polar plumes appear brighter than the surrounding media, suggesting that they
are denser than the background corona or inter-plume regions (e.g. van de Hulst 1950; Saito 1958 \&
1965; etc). The coronal-background electron density is about $\sim10^8$ cm$^{-3}$ and falls sharply
with height (see different density models in Raouafi \& Solanki 2006). In comparison, the density
of polar plumes ranges from $10^8-10^{10}$ cm$^{-3}$ (Del~Zanna et al. 1997; Young et al. 1999).
Young et al. (1999) reported electron densities corresponding to the strong brightenings at plume
foot-points of the order of $2.5-5.6~10^9$ cm$^{-3}$ and found no decrease with height up to 70 Mm.
They also inferred temperatures of the order of 2 MK. In a recent publication, Wilhelm (2006) using
off-limb spectroscopic measurements from the spectrometer SUMER (Wilhelm et al. 1995) on board SOHO
(Domingo et al. 1995) found that plumes are approximately 5-6 times denser than inter-plume regions
up to $\approx1.2-1.3~R_\sun$. The plume-interplume density ratio changes little in this height range.
Note that these are off-limb measurements and they do not take into account the plume area
projected on the solar disk.

The contribution of polar plumes to the fast solar wind arising from the polar coronal holes has
been a subject of debate and controversy. In general, plumes are thought to harbor smaller outflow
speeds than inter-plume regions (e.g., Habbal 1992; Wang 1994;  Hassler et al. 1999; etc). Thus,
polar plumes observed above the solar limb seem to be nearly in hydrostatic equilibrium and show
small or no line shifts, suggesting that the bulk of the acceleration of the solar wind takes place
at larger heights (see Del~Zanna et al. 1997; Kohl et al. 1997; Wilhelm et al. 2000; Teriaca et al.
2003). Doppler dimming techniques applied to SOHO/UVCS (Kohl et al. 1995) observations of the
\ion{O}{6} lines have shown outflows in inter-plume dark lanes ranging from 100 to 150 km~s$^{-1}$
at $1.7~R_{\sun}$ from Sun center and smaller speeds in bright plumes (0-65 km~s$^{-1}$; see
Giordano et al. 2000). However, Gabriel et al. (2003) used the Doppler dimming technique applied to
SOHO/SUMER data and found the intensity ratio of the \ion{O}{6} doublet to be smaller in plumes
than in inter-plumes. They concluded that outflow speeds of the solar wind are greater in plumes
than in inter-plumes. Unfortunately, they did not analyze line profiles, in particular the line
widths, on which the Doppler dimming effect depends.

EUV coronal emissions in spectral lines that are formed within narrow temperature ranges are
excellent temperature diagnostics for different coronal structures. SOHO/EIT (Delaboudini\`ere et
al. 1995) images (\ion{Fe}{12} 195~{\AA} that forms at $\sim1.6$ MK and
\ion{Fe}{9}/$\!\!$\ion{}{10} 171~{\AA} at $\sim1.0$ MK) reveal that plumes are cooler than the
background corona (DeForest et al. 1997; Young et al. 1999). Spectral analysis have also shown that
UV lines are narrower in plumes than in inter-plume regions (e.g. \ion{O}{6} 1032~{\AA} line widths
are 10-15\% lower; Hassler et al. 1997; etc). This suggests that the plume material is relatively
cooler than or has less turbulence than dark lane plasma (inter-plume regions). Wilhelm et al.
(1998) used intensity ratios of temperature sensitive lines for accurate measurements of
temperatures in the low corona. They found evidence for an increasing temperature with height in
the background corona, but no rise of temperature in plumes (see also Young et al. 1999).

The geometry of plumes is an important key in understanding their contribution to the fast solar
wind. The high-speed solar wind at even moderate solar latitudes is thought to originate within
coronal holes at high latitudes and to be affected by superradial expansion of the coronal hole
magnetic field (Zirker 1977). From white light observations plumes are thought to also expand
superradially with altitude (Ahmad \& Withbroe 1977; Munro \& Jackson 1977; Fisher \& Guhathakurta
1995). DeForest et al. (1997) found that plumes expand highly superradially (with a half-cone angle
of 45\degr) in their lowest 20-30 Mm, and more slowly above. DeForest et al. (2001b) also found
that below $\sim5~R_{\sun}$ plumes expand superradially at the same rate as the background coronal
hole. However, Woo \&\ Habbal (1999) and Woo et al. (1999) have claimed to observe radial, rather
than superradial, expansion of plumes in the coronal holes. Speculation on interaction between
plume and inter-plume material has been reported, in particular at high altitudes in the corona.

Saito and Tanaka (1957a,b) were the first to associate plumes with magnetic flux concentrations
(faculae) in the polar regions. Polar plumes arise from footpoints 2-4 Mm ($\approx4\arcsec$) wide
and expand rapidly with height (see DeForest et al. 1997). A simulation has shown that a typical
plume width at the base of the corona is 20 to 28 Mm (Klaus Wilhelm, private communication).
Several authors (Harvey 1965; Newkirk \& Harvey 1968; Lindblom 1990; Allen 1994; etc.) have
suggested that polar plumes arise mainly from unipolar magnetic structures coinciding with the
supergranular network boundaries. However, DeForest et al. (1997) have shown that not all of the
flux concentrations give rise to polar plumes. Other authors suggest that unipolar magnetic flux
concentrations reconnect with ephemeral emergent flux having opposite polarity (Saito 1965; Wang
1994, 1998; Wang \& Sheeley 1995; etc). This is supported by the high temperature observed at the
plume footpoints, producing high pressure plasma from the dipole that can escape along open field
lines. DeForest et al. (2001b) found that polar plumes are of episodic nature: ``polar plumes are
both transient and persistent: they are recurring structures that brighten for only 1 day but
reappear at approximately the same location for up to 2-3 weeks''. They speculate that this
behavior is the result of multiple magnetic reconnections of the unipolar flux concentrations at
the network boundaries encountering multiple ephemeral dipoles driven by supergranular motion.

We consider model plumes having different outflow speeds and turbulence and calculate the line
profile shape of the plume plasma. We use the observed profile shapes of the \ion{O}{6} lines as
primary criteria to select the appropriate height variation of those quantities. We then refine them
to reproduce the other measured quantities, i.e. line widths, total intensities and intensity
ratios. In the following Sections, we study the effect of polar plume locations along the line of
sight on the line profiles and intensities of the \ion{O}{6} doublet and Ly$\alpha$. The comparison
with the observed quantities deduced from SOHO/UVCS spectra allows us to constrain the dynamic
properties of the plasma inside plumes. In section 2, we present a summary of plume spectroscopic
observations by SOHO/UVCS and highlight the spectral properties of \ion{O}{6} 1032~{\AA}. A polar
plume model is described in section 3 and the results from this model are presented and discussed in
sections 4-7. Conclusions and prospects are to be found in section 8. A preliminary version of some of
the results has been reported by Raouafi et al. (2006).

\section{Summary of UVCS plume observations}

The UltraViolet Coronagraph Spectrometer (UVCS; Kohl et al. 1995) has provided important spectral
information on coronal lines in polar holes. The main spectral lines recorded by UVCS are
Ly$\alpha$, the \ion{O}{6} doublet at 103~nm and the \ion{Mg}{10} doublet centered at around 61~nm
(in the order of decreasing emission).

Spectral analysis of UVCS data has shown that \ion{H}{1}~Ly$\alpha$ has, to a good approximation, a
Gaussian shape at all heights accessible to this instrument. Kohl et al. (1999) pointed out small
deviations from Gaussians at the profile peaks, allowing for two line components resulting from the
contributions of different plasma media with different properties along the LOS. Giordano et al.
(1997) have also shown that Ly$\alpha$ profiles are narrower when ray paths intersect high density
structures (plumes) than otherwise (inter-plumes).

The change with height of the \ion{O}{6} line profiles is more complex than for Ly$\alpha$. While
at low altitudes (say below $\sim1.35~R_{\sun}$) the observed profiles are relatively Gaussian
shaped, with small deviations in the wings and peaks, more pronounced deviations of the observed
profiles from a single Gaussian are clearly seen in the observations at higher altitudes up to
about $2.0~R_{\sun}$. At this height, the \ion{O}{6} profiles are well represented by two Gaussian
components: a narrow and a broad one. The latter appears to be associated with regions of low
density (inter-plume regions) while the narrow component shows a clear association with the denser
regions along the LOS (higher emission regions: e.g., polar plumes). The spectral behavior of the
two components as a function of height above the limb is quite different. The width of the broad
component increases dramatically  with height, while that of the narrow one  remains almost
constant (see Kohl et al. 1997; 1999). \ion{Mg}{10} lines have a narrow component at
$r=1.34~R_{\sun}$, accounting for only a small fraction of the observed spectral radiance. For
higher altitudes, the observations are not good enough to distinguish whether or not a narrow
coronal component exists. Kohl et al. (1999) have shown also that the $v_{1/e}$ values for the
narrow components of \ion{O}{6} and \ion{Mg}{10} are smaller than those for \ion{H}{1} 1216~{\AA}
at the same heights, and the narrow \ion{Mg}{10} value is smaller than that of \ion{O}{6}.

At the solar activity minimum the polar coronal holes are well developed and bordering streamers
appear not to reach the UVCS observation altitudes, making it very likely that narrow components are
contributions of fine structures inside the polar coronal hole. Polar plumes are the most prominent
features at the heights at which UVCS observes, which makes them the main source of the narrow component
observed in the spectral lines.

\begin{figure*}[!t]
\centering
\plotone{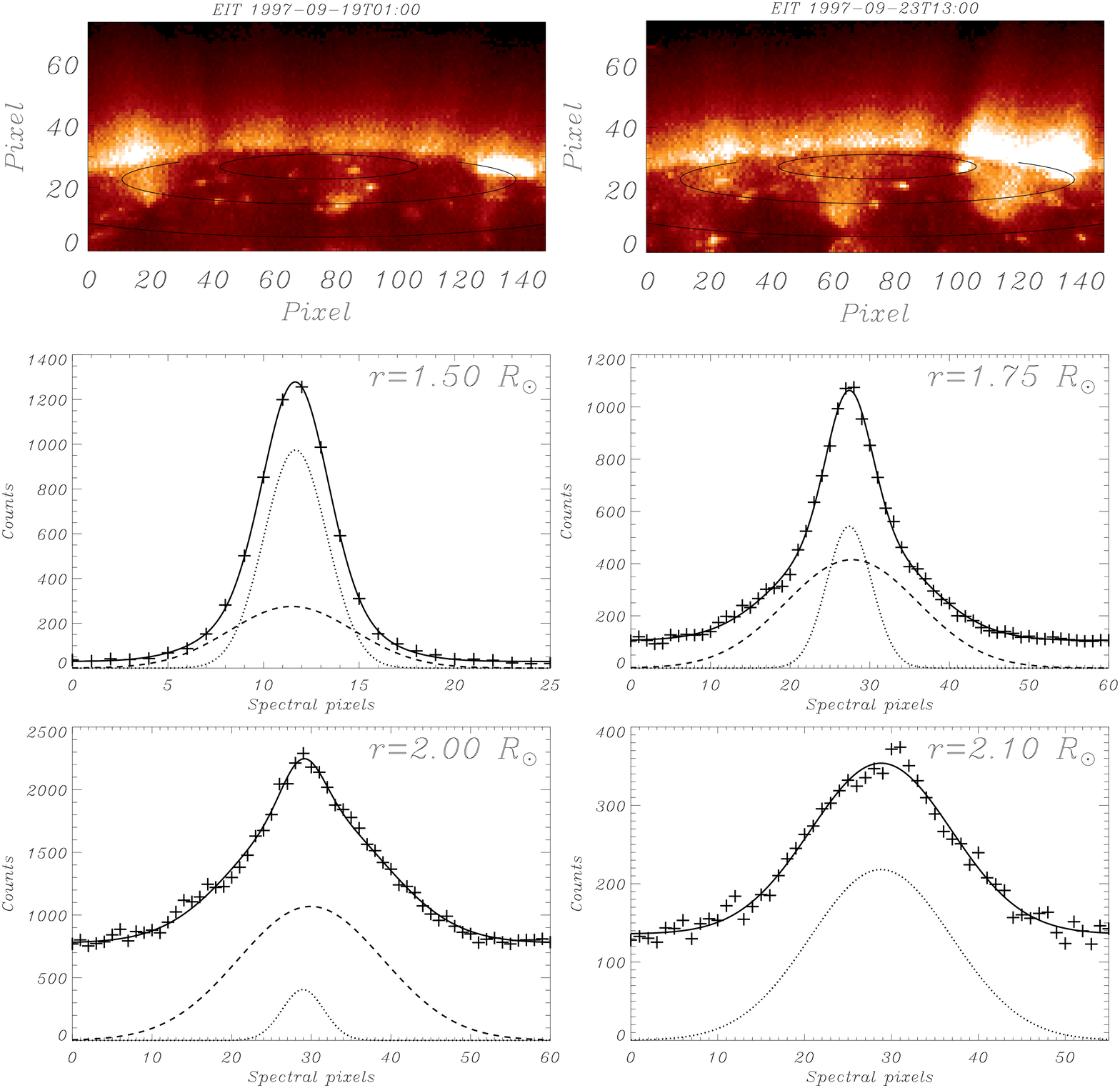}
\caption{Top: EIT 171~{\AA} images of 1997 Sep. 19 and 23, respectively. Polar plumes with strong
emissions are present more than $10\degr$ away from the pole on both days (latitude marked every
$5\degr$). Middle \& bottom: \ion{O}{6} 103.2 nm line profiles ($+$ signs) recorded by UVCS over the
same period of time. The profiles are better fit by two Gaussians (solid curves) except for the
bottom-right panel where a one-Gaussian fit is equally good. The narrow and wide fit components are
represented by dotted and dashed lines. Counts in the different panels are not normalized to
exposure times.}
\label{september19-231997}
\end{figure*}

\begin{figure*}[!t]
\centering
\plotone{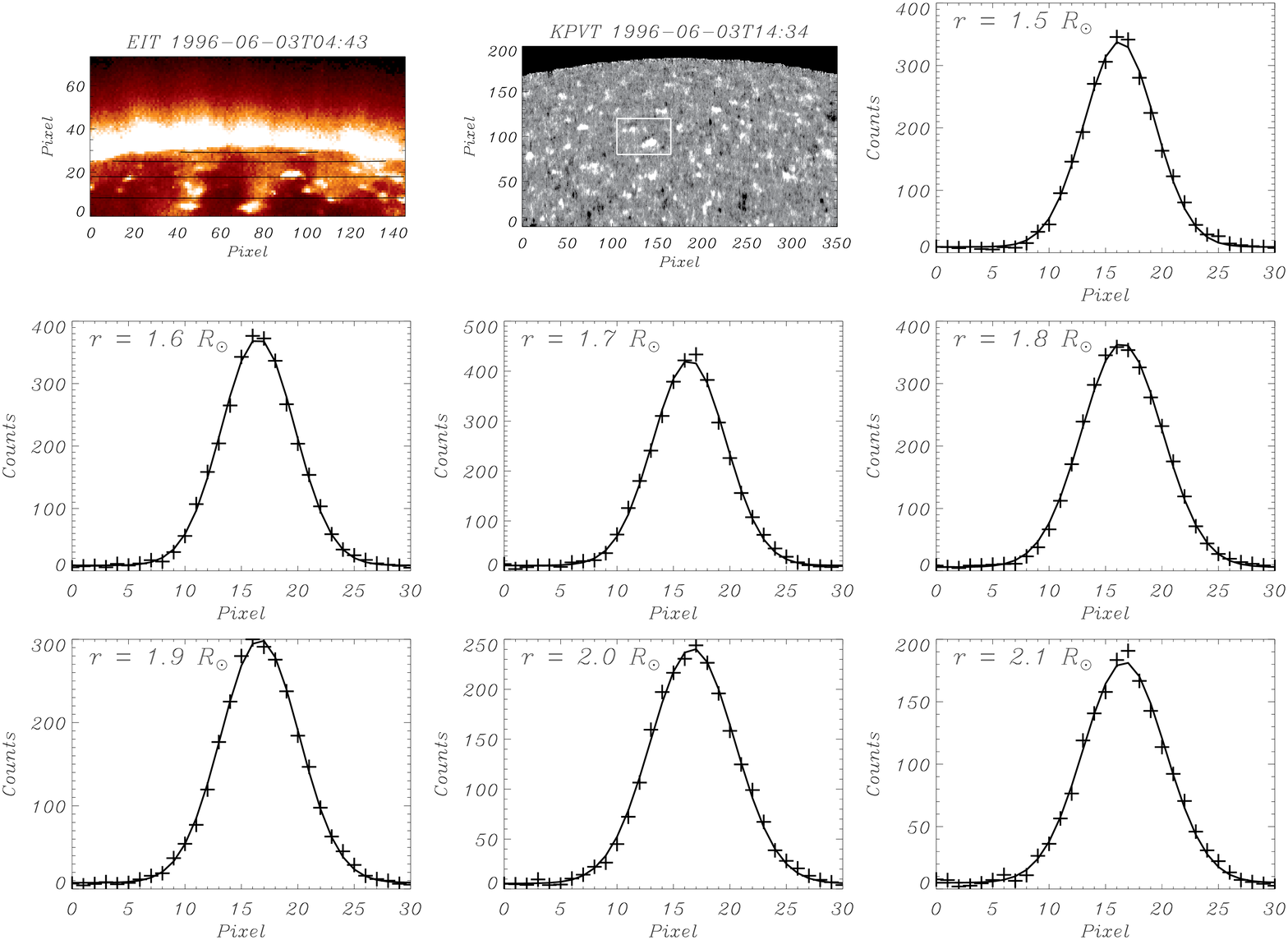}
\caption{EIT 171~{\AA} image (top-left) and Kitt Peak vacuum telescope magnetogram (top-middle) of the
north solar pole on June 3, 1996. Latitude lines indicate $5\degr$ increments. Polar plumes with
footpoints about $20\degr$ away from the pole are clearly visible in the EIT image. The footpoint of
the plumes around pixels $45\times10$ in the top-left panel lies in the white box in the
magnetogram. The other panels display the line profile of \ion{O}{6} 103.2 nm recorded by UVCS on
the same date at different heights (+ signs). The profiles are slightly asymmetric although they are
fit relatively well by a single Gaussian (solid curves).}
\label{june031996}
\end{figure*}

Figs.~\ref{september19-231997} and \ref{june031996} display spectral profiles of \ion{O}{6} 1032
{\AA} recorded directly above the pole on 1997 September 19-23 and 1996 June 03, respectively,
using the $\approx40\arcmin$ long UVCS slit. EIT images and Kitt Peak Vacuum Telescope magnetograms
clearly show the presence of polar plumes against the solar disk and magnetic concentrations with
footpoints at $10\degr-15\degr$ (Fig.~\ref{september19-231997}) and $\approx20\degr$
(Fig.~\ref{june031996}) away from the north pole.  The profiles in the first figure are formed of
two distinct components (narrow and broad). Two Gaussians are needed to give satisfactory fits to
the profiles at $r=1.75~R_\sun$ and partly also around $r=2.0~R_\sun$, although with a small
contribution of the narrow component. However, the profiles in the second figure are very close to
a Gaussian in shape, although they show a small degree of asymmetry.  Thus, the spectral profiles
depend relatively strongly on the distance of the plume footpoints to the pole and also on to the
brightness of the plume. In other words, they depend on where the plume crosses the LOS at a given
altitude. This difference in the profile shapes is a valuable tool to study plasma dynamic in these
coronal fine structures. 

The characteristics of the plumes' spectral contribution (narrow component) to \ion{O}{6} lines as
observed by UVCS are summarized as follows:
\begin{itemize} 
\item Dominates the profiles below $\sim2.0~R_{\odot}$ and decreases above;
\item Not seen in the profiles beyond $\sim2.5~R_{\odot}$. Only the broad
component remains above this altitude. Note that the noise
level in the profiles increases with height, which might screen small contributions from fine
coronal structures along the LOS ; 
\item Mostly located at the center of the broad component which is close to
the rest wavelength of the spectral lines. Thus, no significant Doppler shifts are observed for the
narrow component;
\item The width of narrow component hardly changes as function of height in the range where this
latter is observed (see Kohl et al. 1997).
\end{itemize} 

\section{A Model of polar plumes}

\begin{figure}
\plotone{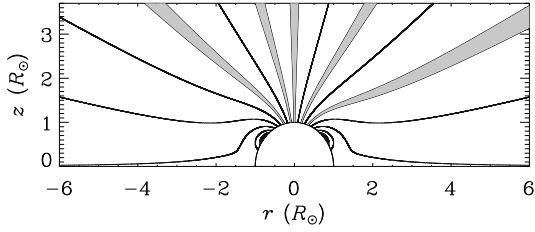}
\caption{Spatial distribution of the polar plumes used for the calculations of the coronal line
   profiles in the plane of the LOS. The plume footpoints are 18 Mm wide and are centered at angular
   positions relative to the pole from left to right $-15\degr$, $-5\degr$, $0\degr$, $10\degr$ and
   $18\degr$, respectively. The magnetic field is assumed to expand similarly in both plume and
   inter-plume regions. No plumes are placed further than $20\degr$ since these plumes do not reach
   large apparent heights above the pole due to the rapid expansion of the magnetic field. However,
   their presence may affect profiles at low altitude (below $\approx2.5~R_{\sun}$). 
   \label{PPmes_Spat_Dist}}
\end{figure}

\begin{figure}
\epsscale{.80}
\plotone{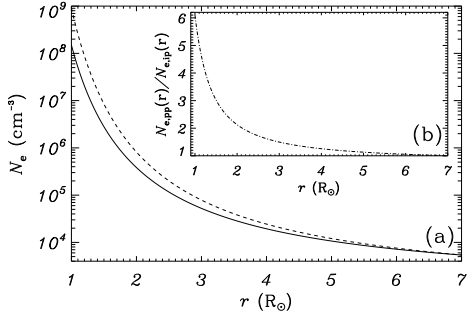}
\caption{(a): Electron density stratifications inside plumes (dashed line) and inter-plumes (solid
 line). The inter-plume density is that obtained from SOHO (SUMER, UVCS and LASCO) data (see Doyle
 et al. 1999a,b). The plume density stratification is obtained by imposing a greater density at the
 solar surface (see text for reference) and lower temperature in plumes relative to the surrounding
 ambient corona. (b): Plume-interplumes density ratio as function of height.}
 \label{IPPP_densities}
\end{figure}

In order to study the influence of polar plumes on the LOS-integrated profiles of \ion{O}{6} and
\ion{H}{1}, we consider a number of plumes aligned along the LOS, with foot points placed at
different angular locations in the polar coronal hole (see Fig.~\ref{PPmes_Spat_Dist}). The
importance of the angular locations of the plume footpoints is enhanced because of the superradial
expansion of the magnetic field. The footpoint location determines the LOS outflow speed of the
plume plasma at the location where the plume crosses the LOS. Based on the work of DeForest et al.
(2001b) and others we assume that plumes expand superradially at the same rate as the surrounding
background corona (inter-plumes) according to the global magnetic model by Banaszkiewicz et al.
(1998). The electron density is considered to be greater in plumes than in inter-plumes, following
Wilhelm et al. (1998) and Wilhelm (2006). The density contrast is largest just above the solar
surface and decreases gently with height, such that it disappears above $\approx7~R_{\sun}$ (see
set in Fig.~\ref{IPPP_densities}). The inter-plume electron density stratification is obtained from
SOHO data (SUMER, UVCS and LASCO (Brueckner et al. 1995); see Doyle et al. 1999a,b: solid line in
Fig.~\ref{IPPP_densities}). The density profile of plumes (dashed line in
Fig.~\ref{IPPP_densities}) is obtained by considering a higher density at the footpoints at the
solar surface ($\approx9\times10^8~\mbox{cm}^{-3}$; see Del~Zanna et al. 1997; Young et al. 1999)
and a lower temperature inside plumes ($\approx10^6~\mbox{K}$) than in inter-plumes
($1.5\times10^8~\mbox{cm}^{-3}$ and $1.2\times10^6~\mbox{K}$, respectively, for the model by Doyle
et al. 1999a,b). The variation of both densities as a function of height is given by Eq.~(1) in
Doyle et al. (1999a,b).

We assume that the considered coronal species have simple Maxwellian velocity distributions inside
both plume and inter-plume regions, but with different velocity turbulence values, $\alpha_s(r)$,
that are functions only of height in both structures. The plume outflow speed, $V_{pp}$, is assumed
to be proportional to the solar wind speed in inter-plume regions, $V_{ip}$, and the proportionality
coefficient is assumed to depend only on the radial distance to Sun center: $V_{pp}(r)=g(r)\times
V_{ip}(r)$, where $V_{ip}(r)$ is obtained through the mass-flux conservation equation. Because of
the lack of conclusive observational results concerning $g(r)$ and $\alpha_{s,pp}(r)$ we consider
height profiles for each. 

Two cases are considered for the plume velocity turbulence: $\alpha_{s,pp}(r)=$ constant
($\approx50~\mbox{km~s}^{-1}$ for the \ion{O}{6} lines following observational values from Kohl et
al. 1997, 1999) and a height-dependent turbulent velocity $\alpha_{s,pp}(r)=f(r)$. The value of
$50~\mbox{km~s}^{-1}$ is approximately the observed width of the narrow component as measured at
$1.34~R_\sun$ (see Fig.~3 of Kohl et al., 1999). In the latter case a low value of the turbulence
is prescribed at low altitudes, which increases to reach the inter-plume value above
$r\approx6.0~R_\sun$. Three cases are considered for $g(r)$: constant, linear and non-linear.

\section{Test cases of the turbulence and outflow evolution inside polar plumes}

The data presented in Figs.~\ref{september19-231997} and \ref{june031996} correspond to plumes based
between $\approx10\degr-20\degr$ from the pole. In order to compare our simulations with these
observed profiles, we limit ourselves in this section to the very left plume in
Fig.~\ref{PPmes_Spat_Dist} that has its foot-point centered at $15\degr$ from the pole. This plume
lies between the examples shown in Figs.~\ref{september19-231997} and \ref{june031996}.

In the present section, we consider the four test cases (sections 4.1-4.4) presented in Table~1.
Note that we have computed the lines also for the remaining combinations of $\alpha_{s,pp}$ and
$V_{pp}$, but nothing new was to be learned from them.

\begin{deluxetable}{ccc}
\tablecolumns{2}
\tablewidth{0pc}
\tablecaption{Test cases of the turbulence and outflow height dependence in polar coronal plumes.
$C$, $C^\prime$, $a$ and $b$ are constants.}
\tablehead{
\colhead{Case}  & \colhead{$\alpha_{s,pp}(r)$} & \colhead{$V_{pp}(r)$}}
\startdata
1.   & $C$    & $C^\prime\times V_{ip}(r)$ \\
2.    & $C$    & $\left\{
\begin{array}{ll}
(a\;r+b)\times\; V_{ip}(r) & \mbox{ if }V_{pp}<V_{ip}; \\
 V_{ip}(r) & \mbox{ otherwise}\\
  \end{array}
  \right.$   \\
3. & $f(r)$ & $\left\{
\begin{array}{ll}
(a\;r+b)\times\; V_{ip}(r) & \mbox{ if }V_{pp}<V_{ip}; \\
 V_{ip}(r) & \mbox{ otherwise}\\
  \end{array}
  \right.$  \\
4. & $f(r)$ & $g(r)\times V_{ip}(r)$ \\
\enddata
\end{deluxetable}

For the inter-plume regions, the plasma properties are similar to those found by Raouafi \& Solanki
(2004 \& 2006). The theoretical formalism used here for the line profile computations is discussed
by Raouafi \& Solanki (2004) and references therein. In the present section, we limit ourselves to
discussing the profiles of the \ion{O}{6} 1032~{\AA} line. \ion{O}{6} 1037~{\AA} is also computed
and is found to behave similarly to the 1032~{\AA} line. Ly$\alpha$ profiles are also computed, but
this line is less sensitive to the behavior of $\alpha_{s}$ and $v$ and its shape does not change
as much as that of \ion{O}{6} 1032~{\AA}. Thus, in contrast to the \ion{O}{6} lines, the Ly$\alpha$
profile is close to a Gaussian for most plume parameters and at most heights.

\subsection{Case 1: Constant velocity turbulence and proportional flow} 

In a first step, we consider a plasma outflow speed inside plumes that is equal to a constant
fraction of that in the inter-plume regions. Plume speeds that are smaller and greater than
inter-plume values are considered ($V_{pp}(r)/V_{ip}(r)=0.25,~0.5,~1.0$ \& $1.5$). $\alpha_{s,pp}$
is height independent and equal to 50~km~s$^{-1}$.

\begin{figure*}[!t]
\centering
\epsscale{1.}
\plotone{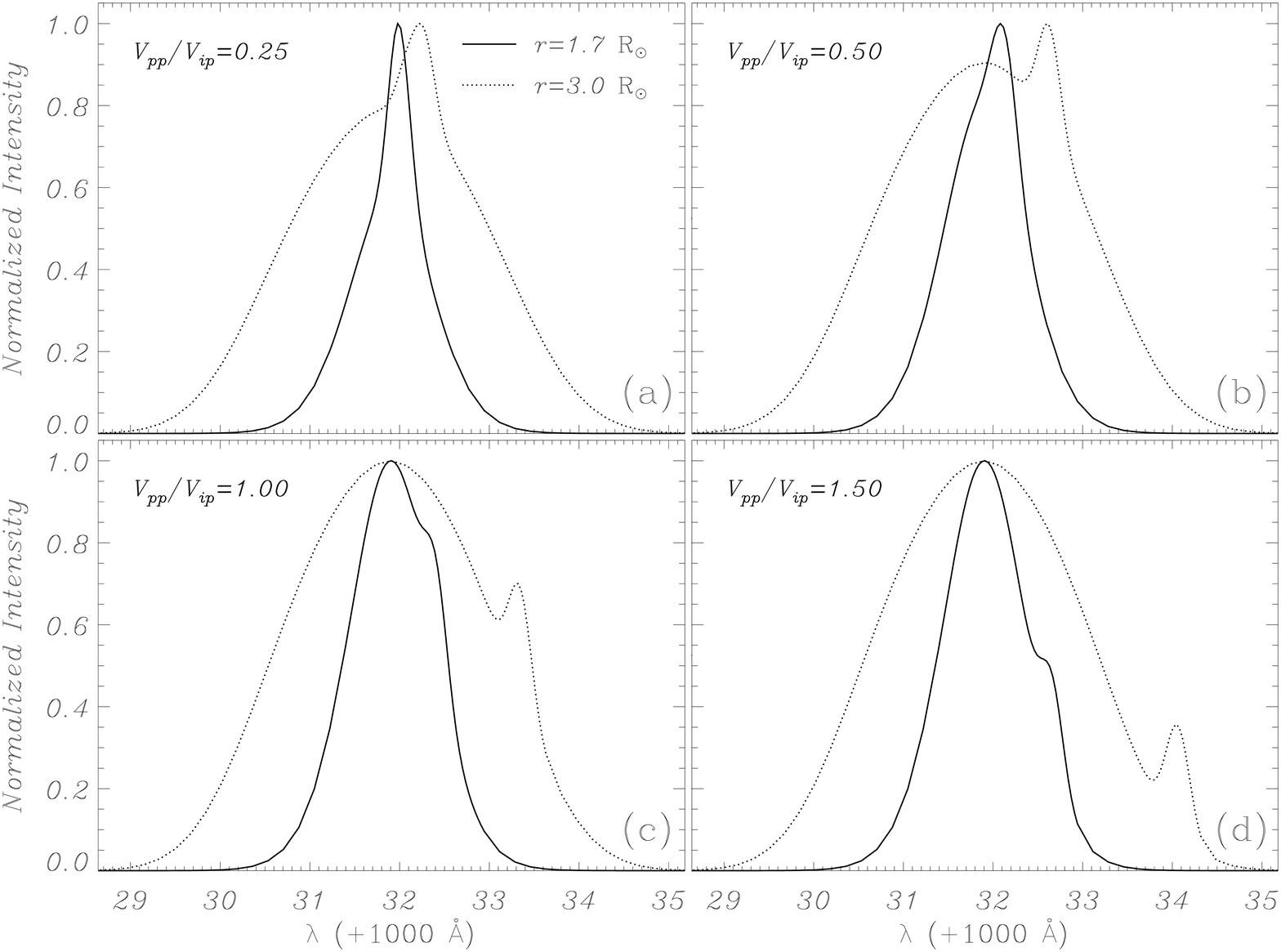}
\caption{Case 1: LOS-integrated model profiles of the \ion{O}{6} 1032~{\AA} line at
   $1.7~R_{\sun}$ (solid lines) and $3.0~R_{\sun}$ (dotted lines) in the presence
   of the very left polar plume in Fig.~\ref{PPmes_Spat_Dist}. The plume
   velocity turbulence is assumed to be constant and the outflow speed is equal
   to a height-independent fraction of the inter-plume outflow speed as noted in
   the different panels.} \label{Fig_O6_pp_pap_ASC_VppCVip}
\end{figure*}

Fig.~\ref{Fig_O6_pp_pap_ASC_VppCVip} displays the profiles of the LOS-integrated \ion{O}{6}
1032~{\AA} line at $1.7~R_{\sun}$ and $3.0~R_{\sun}$ for different $V_{pp}/V_{ip}$ values. For small
$V_{pp}/V_{ip}$, the plume contribution (narrow component) dominates the profiles at low altitudes,
showing some resemblance with the observed profiles (see solid lines in
Figs.~\ref{Fig_O6_pp_pap_ASC_VppCVip}a-b). However, at high altitudes the narrow component is still
clearly present with a strength well above the typical noise level present in the observed profiles.
For greater values of $V_{pp}/V_{ip}$ (Figs.~\ref{Fig_O6_pp_pap_ASC_VppCVip}c-d), the plume
contribution is Doppler shifted away from the line peak already at low altitudes (due to the
location of the considered plume relatively far from the pole). This produces strongly asymmetric
profiles, unlike the observed ones. Also, the plume contribution is still significant at high
altitudes. The present case ($\alpha_{s,pp}=\mbox{constant \& }V_{pp}/V_{ip}=\mbox{constant}$) does
not reproduce the overall observed profiles. Consequently, it does not reflect the plasma conditions
and mechanisms and taking place in polar plumes.

Nonetheless, the computations suggest that outflow speeds in plumes are likely to be much smaller
($V_{pp}/V_{ip}<0.25$) than inter-plume values at low altitudes. This avoids large contributions
from plumes far from the central wavelength of the observed line irrespective of their footpoint
location. Plume outflow speeds close to or greater than than the inter-plume values at low altitudes
will be excluded in the rest of the analysis.

\subsection{Case 2: Constant velocity turbulence and linear-height-proportional flow}

In this section, we keep a height-independent velocity turbulence inside plumes
($\alpha_{s,pp}=50$~km~s$^{-1}$). However, the plume-interplume outflow speed ratio is now allowed
to vary linearly with height, such that
\begin{eqnarray}
V_{pp}(r)=\left\{
\begin{array}{cl}
(a\;r+b)\times V_{ip}(r) & \mbox{if $V_{pp}(r)<V_{ip}(r)$,} \\
V_{ip}(r) & \mbox{otherwise,}
\end{array}
\right.
\end{eqnarray}
where $a$ and $b$ are constants chosen such that $(a\;r+b)$ is zero or very close to zero at the
solar surface and increases toward unity at high altitudes, roughly above $6~R_{\sun}$. A possible
mechanism for such an acceleration of the plume plasma could be interaction with the
surrounding background medium having higher outflow speeds.

\begin{figure*}[!t]
\centering
\epsscale{.65}
\plotone{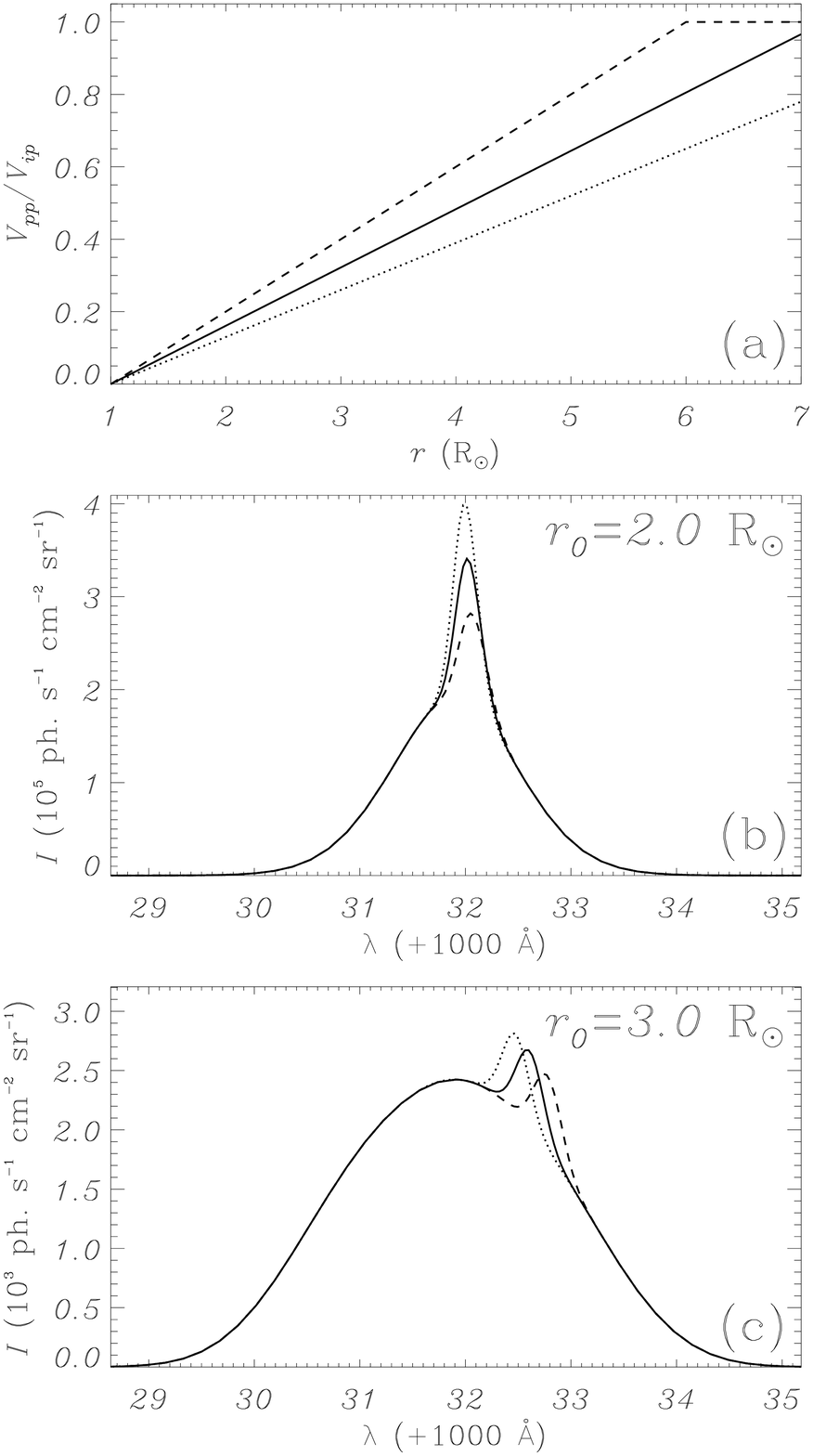}
\caption{Case 2: \ion{O}{6} 1032~{\AA} line profiles for a wind speed in plumes equal to a
    linear-height-dependent fraction of the inter-plume outflow speed and a constant velocity
    turbulence in plumes. (a) The chosen flow speed height profiles, (b) resulting line profiles at
    $2.0~R_{\sun}$ and (c) at $3.0~R_{\sun}$ from Sun center.}
\label{AsConsVppLVip}
\end{figure*}

We consider the three height profiles of $V_{pp}(r)/V_{ip}(r)$ that are given in
Fig.~\ref{AsConsVppLVip}a. LOS-integrated profiles computed along rays reaching $2.0~R_{\sun}$ and
$3.0~R_{\sun}$ from disk center are plotted in Figs.~\ref{AsConsVppLVip}b-c. The obtained profiles
are to some extent similar to those in Fig.~\ref{Fig_O6_pp_pap_ASC_VppCVip} for
$V_{pp}(r)/V_{ip}(r)=\mbox{constant}$. Although the profiles at low altitudes now all have the
narrow component nearly centered at the broad component, which is close to the observations, those
obtained at high altitudes have the same significant narrow contributions shifted to the wings of
the profile. Such a signal is not observed in the corona.

\subsection{Case 3: non-linear-height-dependent velocity turbulence and linear-height-proportional
flow}

\begin{figure}[!t]
\centering
\epsscale{1.}
\plotone{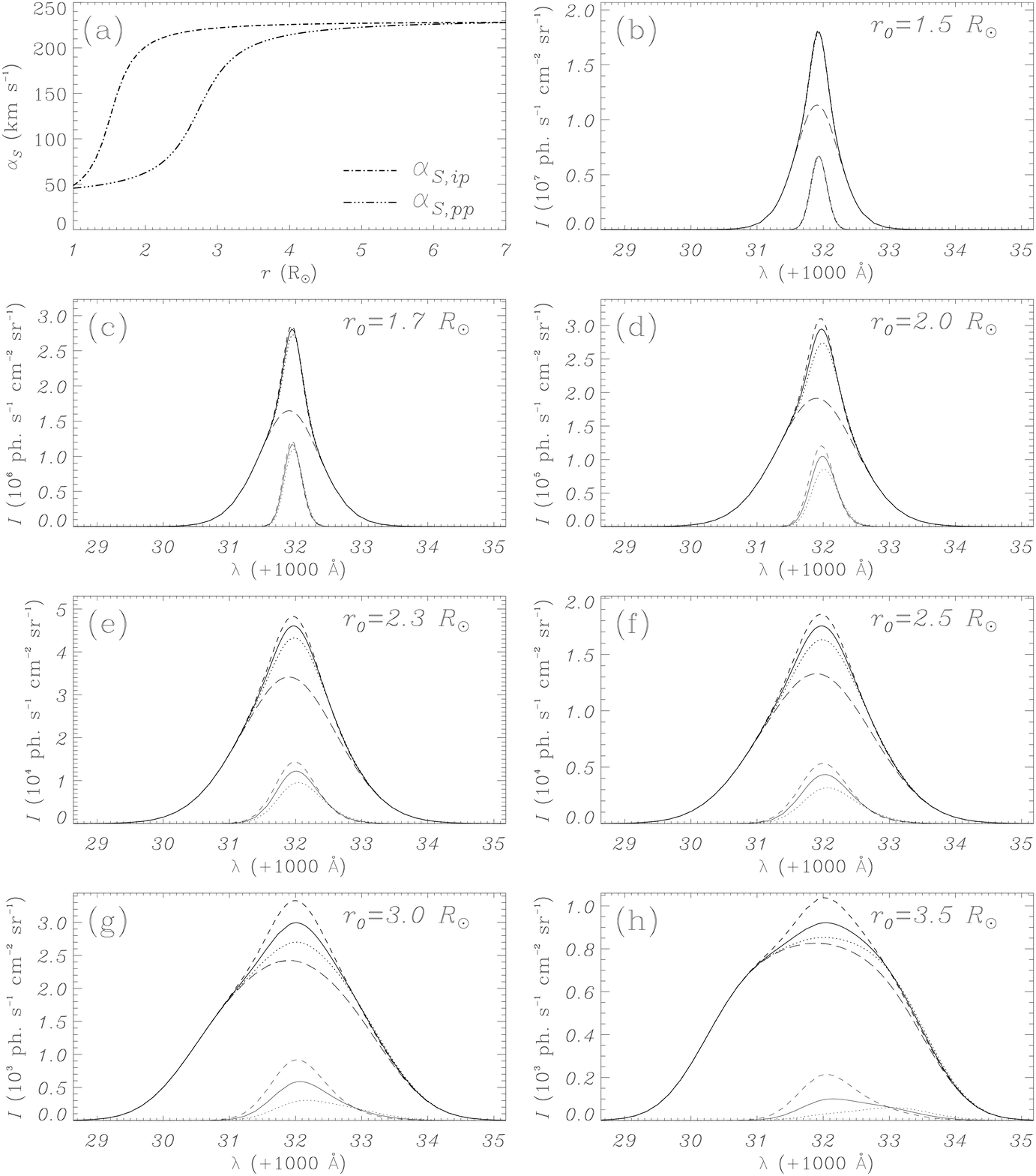}
\caption{Case 3: (a): $\alpha_{s,pp}(r)$ and $\alpha_{s,ip}(r)$ as a function of height used for the
computation of the profiles in frames b-h. (b-h): \ion{O}{6} 1032~{\AA} LOS-integrated profiles
corresponding to the three plume velocity models shown in Fig.~6 are given with the same linestyle
as in that figure. Plume (smaller, narrower profiles at bottom of each frame) and inter-plume
(long-dashed curves) contributions are also plotted.}
\label{VppLVVip_AlpSppNL}
\end{figure}

The plume outflow speed is that of Case 2, while the velocity turbulence of the plume plasma is
assigned a height-dependent function. A variation in the velocity turbulence appears natural if cool
plume and hot inter-plume material interacts. Due to the lack of observational results
concerning the variation of both parameters, we assume, for simplicity, similar profiles for the
turbulence inside plumes and inter-plumes $\alpha_{s,ip}(r)$. Thus, $\alpha_{s,pp}(r)$ has the
following functional form (see Raouafi \& Solanki 2004, 2006)
\begin{eqnarray}
\alpha_{s,pp}(r)=\kappa\times\arctan\left(r^\eta+\mu\right)+\delta,
\label{NL_AlpSpp}
\end{eqnarray}
where $\kappa~(=205/\pi)$, $\eta~(=1.5)$, $\mu~(=4.5)$ and $\delta~(=130)$ are adjustable parameters
whose values have been selected in order to get as close as possible to the observations. Since
there is observational evidence for cooler plume plasma than in inter-plumes, we assume that
$\alpha_{s,pp}(r)$ is smaller than $\alpha_{s,ip}(r)$ at low altitudes and converges
gently towards the large values of $\alpha_{s,ip}(r>6~R_{\sun})$ (see
Fig.~\ref{VppLVVip_AlpSppNL}a). This guarantees the presence of a narrow component at low heights
and progressively broader plume profiles as we move further above the solar limb.

Figs.~\ref{VppLVVip_AlpSppNL}b-h displays the \ion{O}{6} 1032~{\AA} line profiles obtained for
lines-of-sight ranging from $1.5~R_{\sun}$ to $3.5~R_{\sun}$ above the pole for the plume outflow
speed profiles shown in Fig.~\ref{AsConsVppLVip}a. At altitudes below $\approx1.5~R_{\sun}$ the
contribution of polar plumes to the line profile is narrower than that of inter-plumes and is
approximately centered at the rest wavelength of the line. This results in approximately
Gaussian-shaped line profiles that are slightly narrower than those obtained without polar plumes
(see Fig.~\ref{VppLVVip_AlpSppNL}b). For $1.5~R_{\sun}\le r\le2.5~R_{\sun}$ a profile that can be
well represented by two Gaussians and that resembles the observed profiles is obtained
(Figs.~\ref{VppLVVip_AlpSppNL}c-f). The broad component is due to inter-plume material, the narrow
component to plumes. Due to the low plume outflow speeds at these altitudes, the Doppler shift of
the narrow component is low.

Above $\approx2.5~R_{\sun}$, the increase in the plume velocity turbulence results in increasingly
broader plume contributions, so that the line profile (see Figs.~\ref{VppLVVip_AlpSppNL}g-h) takes
on the approximate shape of a single Gaussian for the dashed velocity profile in
Fig.~\ref{AsConsVppLVip}a (dotted lines in the same figure). For the other velocity models, the plume
component is narrower, in particular, for the dotted plume velocity model, so that one can still
distinguish two line components at high altitudes. Hence, high plume outflow speeds at
$r\ge2.5~R_{\sun}$ that are closer to inter-plume values fit the observations better than smaller
ones.

The present case is in better agreement with the observation than cases 1 and 2, although the
agreement is not perfect, both at very low and very high altitudes. E.g. at low altitudes the plume
contribution does not dominate the LOS-integrated profiles. Probably the plume outflow speed at
these altitudes is still high enough to dim the radiative plume component (shifted from resonance
due to the small velocity turbulence). Improvement can be obtained by considering a non-linear
height-dependent $V_{pp}(r)/V_{ip}(r)$ allowing for lower and higher plume speeds at low and high
altitudes, respectively. This is the aim of case 4. 

\subsection{Case 4: non-linear-height-dependent velocity turbulence and
non-linear-height-proportional flow: case best reproducing observations}

\begin{figure*}[!t]
\centering
\epsscale{1.}
  \plotone{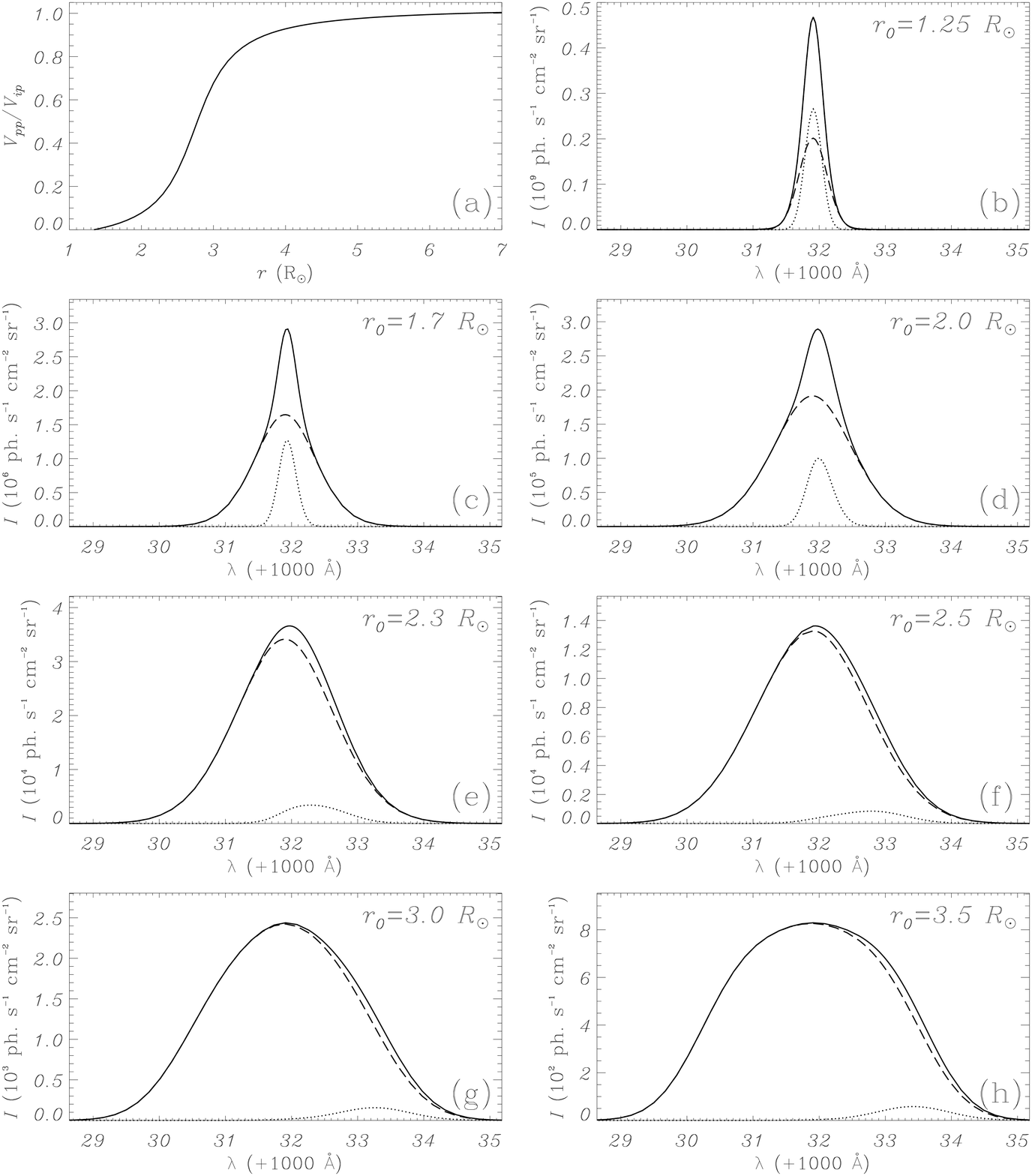}
\caption{Case 4: The same as Fig.~\ref{VppLVVip_AlpSppNL}, but for plume outflow speed equal to a
   non-linear height-dependent fraction of inter-plume speed (a). The line profiles computed along
   different rays now show considerable similarity to the observed ones (b-h). The contribution of
   plumes (dotted profiles) is more significant at low altitudes and is extensively dimmed relative
   to the inter-plume contribution (dashed profiled) at higher altitudes. Line profiles are fit by
   two Gaussians up to 2.0~$R_{\sun}$ and by a single Gaussian for greater heights. 
   \label{VppNLVVip_AlpSppNL}}
\end{figure*}

The velocity turbulence in plumes $\alpha_{s,pp}(r)$ is again described by Eq.(\ref{NL_AlpSpp}) and
is illustrated by the triple-dot-dashed curve in Fig.~\ref{VppLVVip_AlpSppNL}a. The plume outflow
speed is now considered to be related in a non-linear way to the inter-plume speed:
\begin{eqnarray}
V_{pp}(r)=g(r)\times V_{ip}(r),
\end{eqnarray}
where $g(r)$ is illustrated by the solid line in Fig.~\ref{VppNLVVip_AlpSppNL}a. The
plume-inter-plume solar wind speed ratio is less than 10\% below $\approx2.0~R_\sun$ from Sun
center, and increases abruptly to more than 75\% by about $3.0~R_\sun$, then converges gently toward
unity. This profile allows more contribution from the plume radiative component at
$r\le2.0~R_{\sun}$ and favors broader plume profiles at $r\ge3.0~R_{\sun}$.

The LOS-integrated \ion{O}{6} 1032~{\AA} line profiles are displayed in
Figs.~\ref{VppNLVVip_AlpSppNL}b-h, together with the plume (dotted profile) and inter-plume
(long-dashed line) contributions. The plume contribution is now, as expected, more significant at
low altitudes, in particular below $\approx1.7~R_\sun$ and dominates the profile below
$\approx1.4~R_\sun$ (see Fig.~\ref{VppNLVVip_AlpSppNL}b).

The change in the plume contribution is due to combined effects of the electron density decrease but
mostly to the variation of the outflow speed and turbulence of the plasma inside plumes. The
calculated profiles are close to a Gaussian shape at $r\le1.3~R_{\sun}$ and display increasingly a
bi-Gaussian shape above this altitude up to $r\ge2.0~R_{\sun}$ (Figs.~\ref{VppNLVVip_AlpSppNL}c-d).
Above $\approx1.3~R_{\sun}$, the obtained profiles cannot be represented by one Gaussian and the two
components (narrow and broad) can be distinguished even without fitting. The width of the plume
component increases relatively rapidly above $\approx2.0~R_\sun$ and so does its Doppler shift. This
sudden change in the profile of the plume contribution allows the two (broad and narrow) components
to merge completely above $\approx2.3~R_\sun$. The resemblance of the computed profiles to the
observed ones is now uniformly good (compare profiles in Figs.~\ref{september19-231997} \&
\ref{VppNLVVip_AlpSppNL}). Note that greater gas densities in plumes above $7.0~R_{\sun}$ than the
one considered here allow for a greater contribution from plumes that results in broader
LOS-integrated profiles at high altitudes, that are even closer to the observed ones. Any remaining
small differences between computed and observed profiles could also reflect the footpoint location
of the modeled plume, or the possible presence of multiple plumes along the LOS in the observations.

Case 4 best reproduces the observed profile shapes. We expect this model to be the closest to the
plasma conditions in the polar plumes. In other words, the plume material remains cooler and much
slower than the inter-plume plasma up to $\approx2.0~R_\sun$ from Sun center, where a rapid
increase in both outflow speed and effective temperature takes place, leading to plume dynamic
properties converging towards those of inter-plume region within the next solar radius. The rapid
change in plume plasma properties is likely to be due to interaction between the cool and slow
plume and hot and fast inter-plume materials, although some authors speculated that this
interaction takes place much higher in the solar corona (see introduction). In the next sections,
we will consider this case only for an extensive study of the profiles and total intensities of the
different spectral lines considered here. The contribution of polar plumes located very close to or
on the pole will also be discussed.

\section{\ion{O}{6} doublet}

\begin{figure*}[!t]
\centering
\epsscale{.7}
  \plotone{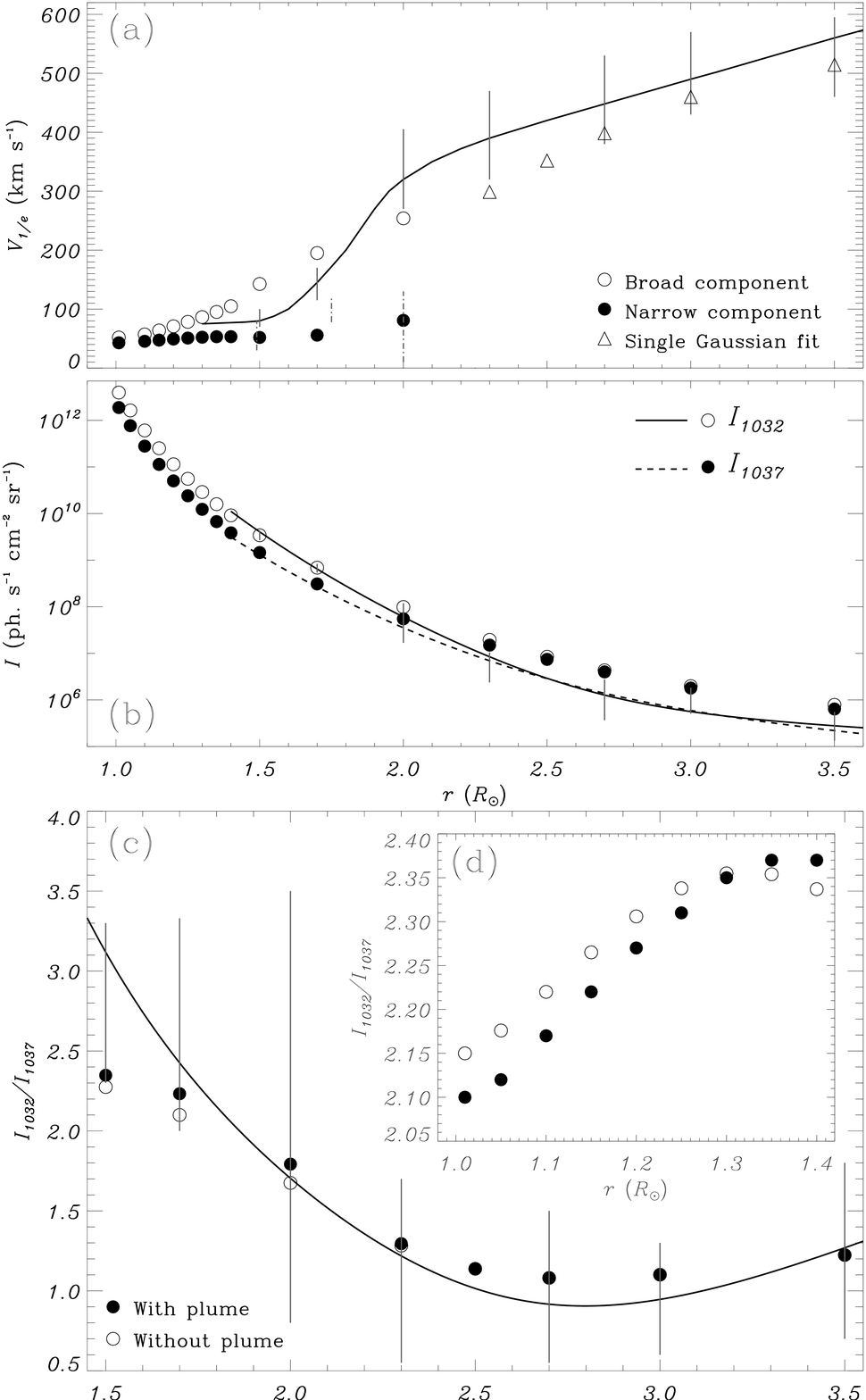}
\caption{Model (circles and triangles) line width (a), total intensity (b) and intensity ratio
   (c,d) of the \ion{O}{6} doublet vs. $r$. Values obtained from synthetic line profiles following
   case 4 (profiles plotted in Fig~8) are shown as filled and open circles and triangles. The solid
   and dashed lines are best fits to UVCS observations, the solid error bars represent the accuracy
   of the observations (see Cranmer et al. 1999). The dot-dashed vertical bars represent the
   uncertainty in the observed narrow component widths (see Kohl et al. 1997).}
   \label{VppNLVVip_AlpSppNL_WTIRat}
\end{figure*}

Fig.~\ref{VppNLVVip_AlpSppNL_WTIRat} displays the height dependence above the pole of the \ion{O}{6}
1032~{\AA} line width (Fig.~\ref{VppNLVVip_AlpSppNL_WTIRat}a) and total intensity
(Fig.~\ref{VppNLVVip_AlpSppNL_WTIRat}b) as well as of the intensity ratio of the \ion{O}{6} doublet
(Figs.~\ref{VppNLVVip_AlpSppNL_WTIRat}c-d) from slightly above the solar limb to $3.5~R_\sun$ from
Sun center. The profiles underlying this figure were computed according to case 4 (section 4.4).
They were fit with two Gaussians below $2.0~R_\sun$ and with one Gaussian above. The widths of the
broad and narrow components and of the single Gaussian fit are represented by different symbols.
Also plotted are values obtained from UVCS spectra together with their corresponding error bars
according to Kohl et al. (1997) and  Cranmer et al. (1999). The narrow component widths hardly
change with height up to $2.0~R_\sun$ and fit the measured values relatively well. The widths of the
broad component, however, change rapidly with height, leading to a non-Gaussian shape of the
composite profiles. Higher up the obtained line widths are comparable to those obtained by Raouafi
\& Solanki (2006; Fig.~8) although they lie somewhat close to the measurements, in particular,
around $2.0~R_\sun$.

Total intensities are also comparable to those computed by Raouafi \& Solanki (2006) although the
intensities below $2.0~R_\sun$ fit the observed values for the \ion{O}{6} 1032~{\AA} line better.
The change in total intensities can be more clearly seen in the intensity ratio of the \ion{O}{6}
doublet. Above $1.5~R_\sun$, the ratios obtained from the composite profiles (full circles) are
improved compared to the case without plumes (open circles), in particular, at low altitudes. All
the computed ratios now lie inside the error bars of UVCS measurements. Note that only simple
Maxwellian velocity distribution are used and no anisotropy in the kinetic temperature of the
\ion{O}{6} ions is considered. Below $1.5~R_\sun$, the intensity ratios with plume contribution
taken into account are smaller than those without plume contribution (see bottom panel of
Fig.~\ref{VppNLVVip_AlpSppNL_WTIRat}). In our computations, this is due to the large electron
density numbers in plumes compared to inter-plume densities and is not a solar wind effect as
reported by Gabriel et al. (2003 \& 2005). In addition, we showed previously that plume wind speeds
faster or just close to inter-plume speeds give line profiles different from those observed in the
polar coronal holes. Here we demonstrate that the fast solar wind originates mainly from inter-plume
regions and that the plume plasma remains slower and cooler up to approximately $2.0-3.0~R_\sun$
from Sun center (see Fig.~9).

\section{\ion{H}{1} Ly$\alpha$ line}

\begin{figure*}[!t]
\centering
\epsscale{1.}
  \plotone{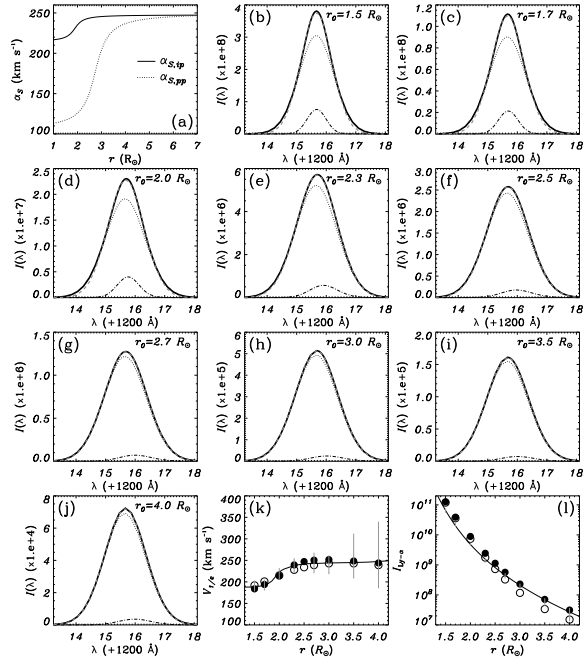}
\caption{(a): velocity turbulence of hydrogen atoms in a plume (dotted line) and in inter-plume
    (solid line) as a function of height. (b-j): computed Ly$\alpha$ profiles: composite profile
    (balck-solid lines); the plume and the inter-plume contributions (dotted and dot-dashed lines,
    respectively) and single Gaussian fits to the composite profiles (gray-dashed lines). Computed
    profile widths (k) and total intensities (l) of Ly$\alpha$ with (full circles) and without
    (open circles; see Raouafi \& Solanki 2006) polar plume crossing the LOS. The solid curves in
    (k) and (l) are the best fits to UVCS measured values, together with the error bars of the
    measurements (see Cranmer et al. 1999).
    \label{LyAlp_PRWTI}}
\end{figure*}

Fig.~\ref{LyAlp_PRWTI}a displays the velocity turbulence used for the computation of Ly$\alpha$
(solid line for inter-plume and dotted line for plume). We assume the same
$v_{pp,Ly\alpha}(r)/V_{ip,Ly\alpha}(r)$ variation of the plume velocity turbulence as in the case
of the \ion{O}{6} ions: small values at low altitudes and a rapid increase toward inter-plume values
above $\approx2.0~R_\sun$. The plume-inter-plume ratio of the outflow speed of the hydrogen atoms is
assumed to be the same as for the \ion{O}{6} ions in the previous section
(see Fig.~\ref{VppNLVVip_AlpSppNL}a). Note that the hydrogen outflow speed in inter-plume regions is
lower than that of \ion{O}{6} ions (see Raouafi \& Solanki 2006).

The computed line profiles of Ly$\alpha$ are shown in Fig.~\ref{LyAlp_PRWTI}b-j (solid-black lines)
together with the plume (dot-dashed lines) and inter-plume (dotted lines) contributions for
different rays crossing the polar solar axis at different heights. The red curves are single
Gaussian fits to the computed profiles. The high quality of the fits at all heights demonstrates how
close the computed profiles are to a Gaussian shape. However, small deviations from the single
Gaussian shape are noticeable, in particular, at low altitudes in the wings and peaks of the
profiles. Kohl et al. (1999) observed similar deviations and interpreted them as due to the
contribution of polar plumes. Note that slightly changing ($\approx10\%-20\%$) the model parameters
(i.e. the outflow speed and turbulence of the plasma inside plumes) does not change much the
obtained profiles, which remain Gaussian shaped to a good approximation.

The widths of the computed line profiles are displayed in Fig.~\ref{LyAlp_PRWTI}k as a function of
the projected heliocentric distance of the corresponding rays (full circles). The widths obtained
without plumes (open circles) are also displayed for comparison, together with the best fit to UVCS
observations (solid curve) and the error bars of the observed values (vertical lines; see Cranmer et
al. 1999). Line widths obtained in both cases (with and without polar plumes) are within the errors
bars of the measured values. However, the widths obtained with plumes taken into account fit the
observed values better. By optimizing the number and position of plumes and their properties, an
even better match to the observations may be obtained.

Fig.~\ref{LyAlp_PRWTI}l shows the total intensity of Ly$\alpha$. The best fit to the observed
intensities along with the error bars are also displayed for comparison. At low altitudes,
intensities obtained with plumes taken into account are very similar to those without plumes and are
slightly greater than the measured ones. However, at higher altitudes intensities obtained with
plumes provide an improved fit to the observations. Generally, the presence of plumes along the LOS
slightly increases the total intensity due to the density enhancement in plumes.

\section{On the foot-point locations of polar plumes}

\begin{figure*}[!t]
\centering
  \plotone{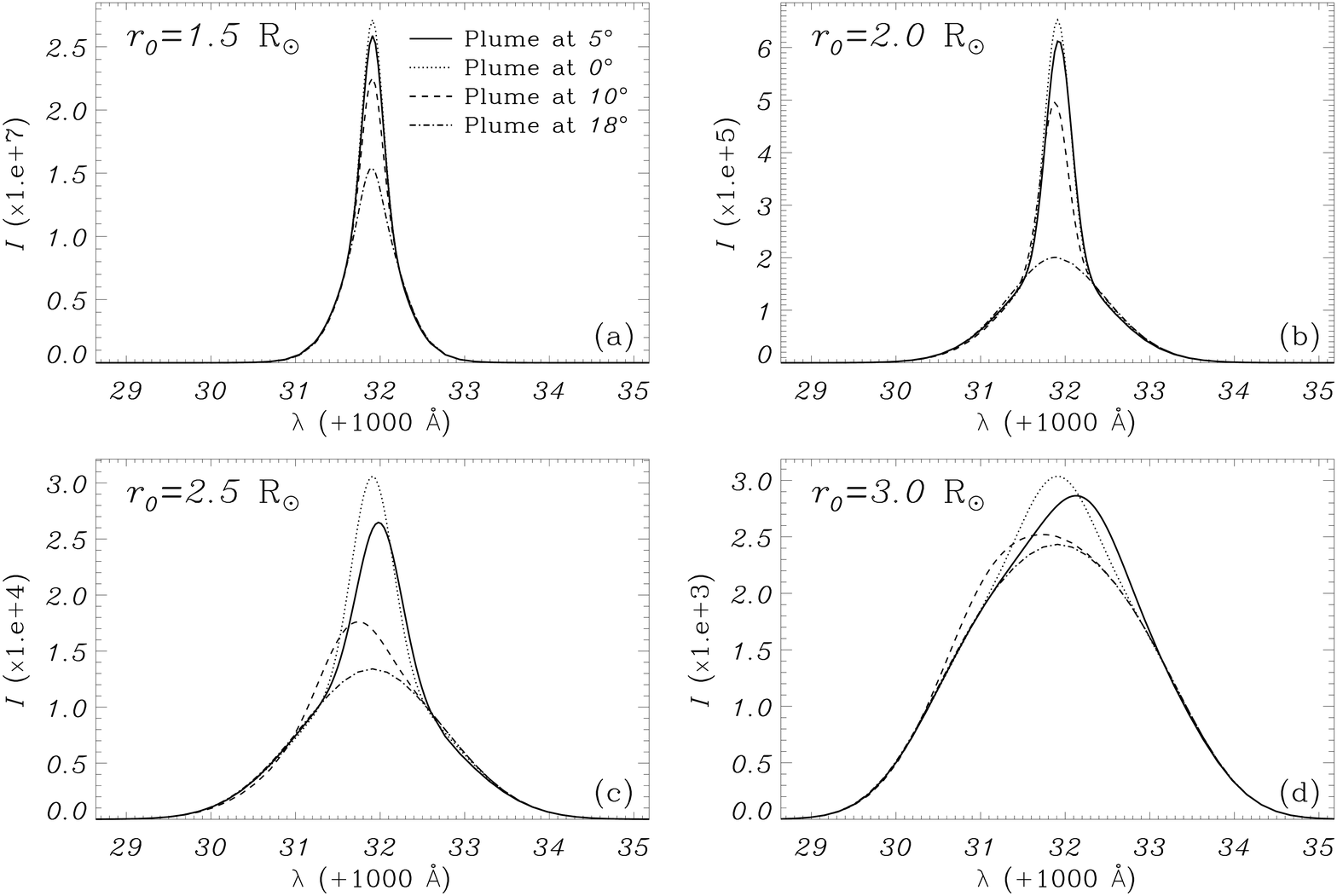}
  \includegraphics[width=1\textwidth]{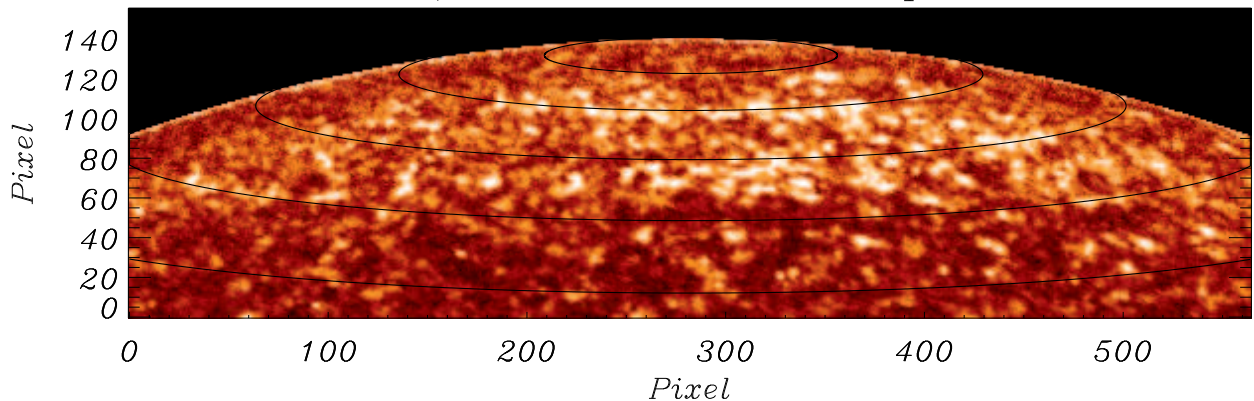}
\caption{(a-d): line profiles of the \ion{O}{6} 1032~{\AA} line at different altitudes with the
    individual contributions of the four right-hand plumes in Fig.~\ref{PPmes_Spat_Dist} taken into
    account separately. The intensity unit is photons s$^{-1}$~cm$^{-2}$~sr$^{-1}$~{\AA}$^{-1}$.
    Contributions from plumes at or close to the pole are noticeable beyond $2.5~R_{\sun}$ which is
    not the case for the observed profiles. Note the great similarity of the dot-dashed profiles
    (with contribution from the farthest plume to the pole ($18\degr$)) with the observed profiles
    in Fig.~\ref{june031996}. Bottom: SOLIS magnetograms averaged over September 2005 showing the
    unsigned flux distribution around the north polar coronal hole. The flux is concentrated
    preferentially between 70\degr and 80\degr latitudes. The solid curves mark latitudes every
    5\degr. } \label{Footpoints_fig}
\end{figure*}

Fig.~\ref{Footpoints_fig}a-d shows the LOS-integrated profiles of the \ion{O}{6} 1032~{\AA} line
with contributions from each of the four right-hand plumes in Fig.~\ref{PPmes_Spat_Dist} taken into
account separately. The contributions from the two polar plumes based at colatitude $-5\degr$ and
$0\degr$ (the narrow component of the line profile) dominate the profiles up to
$\approx2.5~R_{\sun}$ and remain clearly visible up to $\approx3.5~R_{\sun}$ (dotted and solid
lines) which is not the case for the observed profiles. The reason for the continuing visibility of
the truly polar plumes up to large altitudes is that the rapid acceleration of the solar wind in
the plume hardly affects the plume profile, since it is almost perpendicular to the LOS. The
contribution from the plume at $10\degr$ is important at low altitudes but relatively small at high
altitudes (dashed lines). This suggests that polar plumes preferentially originate more than
$\approx10\degr$ away from the pole. The contribution from the plume with footpoint at $18\degr$
(dot-dashed lines) is very similar to that shown in Fig.~\ref{june031996} due to  plumes at
$\approx20\degr$ from the solar pole.

Various authors have studied locations of polar plume foot-points within polar regions. They traced
plumes arising near the edges of coronal holes, relatively far away from the solar pole (eclipse
observations; see Saito 1965). In contrast, we did not find any report on plumes very close to the
polar axis. This case may be very rare for unknown physical reasons (for example, bipolar flux may
not emerge near the pole).

Polar coronal plumes can be easily identified off-limb in white light (LASCO, eclipses) and EUV
(EIT, TRACE) images. On the solar disk, for instance by using EIT images, it is also possible to
trace them back to the solar surface, in particular, using images recorded in the wavelength bands
containing the hot lines 171~{\AA} and 195~{\AA} that are regularly recorded by EIT. X-ray images
also allow the determination of the locations of plume foot-points, that correspond very often to
bright points reflecting the high temperatures there. Once the plume foot-points are identified, it
is relatively easy to link them to the underlying magnetic structures using magnetograms. Data
recorded daily over several consecutive days or better still weeks help to survey the latitudes of
polar plume foot-points. For this purpose, periods of time when the solar axis is tilted toward the
observer are more useful since the observation of a major area of the polar coronal hole including
the pole is possible. These times correspond to March and September of each year.

EIT images recorded at solar minimum seem to support this hypothesis. An average of all daily
unsigned SOLIS (Synoptic Optical Long-term Investigations of the Sun) high sensitivity magnetograms
recorded in September 2005 (see bottom panel of Fig.~\ref{Footpoints_fig}) show magnetic flux
concentrations preferentially between $\approx70\degr$ and $\approx80\degr$ latitude, although we
are not at solar minimum yet. Another reason why plumes are more likely to have their
footpoints some degrees away from the pole  has got to do with the relative area available for
plumes between 70 and $80^\circ$ is 3 times larger than at latitudes above $80^\circ$. However,
preliminary analysis of SOLIS magnetograms tends not to support the area argument. This subject
will be addressed in detail in a subsequent publication.

\section{Discussion}

The aim of the present study is to obtain information on the height evolution of the plasma
dynamics inside polar plumes. We compared profile shapes of the \ion{O}{6} doublet (1032 and
1037~{\AA}) and the \ion{H}{1} Ly$\alpha$ spectral lines observed by SOHO/UVCS at different
altitudes in polar coronal holes to synthetic profiles that include the contribution of polar
plumes placed at different locations along the LOS. Observational constraints on these fine coronal
structures (densities, temperatures, geometry, ...) have been taken into account following
different reports in the literature. Since the outflow speed and kinetic temperature evolution with
height are not conclusively known, we consider different height profiles for the outflow speed and
plasma turbulence inside plumes and study how these choices influence the line profiles of the
\ion{O}{6} doublet, which are more sensitive to the detailed dynamic structure along the LOS than
\ion{H}{1}~Ly$\alpha$.

By comparing the synthetic profiles to the observed ones in the polar coronal holes, we determined
plausible height profiles for the outflow speed and plasma turbulence in polar plumes with respect
to the surrounding background corona. We find that models with constant (height-independent)
velocity turbulence in the plumes show narrow components with a significant contribution to the
LOS-integrated profile at all heights, in particular above $\sim3.0~R_{\sun}$ where there is no
observational indication of the presence of such a narrow component in the observed profiles. This
is independent of whether the plume outflow speed equals a constant or a height-dependent fraction
of the inter-plume outflow speed. Synthetic profiles with shapes very similar to the observed ones
are obtained when the plasma turbulence in plumes is height-dependent with similar variation to
that in inter-plume regions and converges towards the latter at greater heights. At the same time,
the outflow speed is small near the solar surface, but approaches the inter-plume value at greater
heights. The narrow component is found to contribute significantly to the resultant profiles at
altitudes below $\sim2.0~R_{\sun}$ and is increasingly dimmed above that altitude. This model
suggests that the plume plasma remains much slower and cooler (in the sense that
$\alpha_{s,pp}<\alpha_{s,ip}$) than inter-plume material up to $\approx2.0~R_{\sun}$. Above
$\approx2.0~R_{\sun}$ these parameters experience a rapid rise towards inter-plume values reached
by $\approx3.0~R_{\sun}$. The rapid acceleration and increase in the effective temperature of the
plasma inside plumes is probably the result of interaction with the faster and hotter inter-plume
material. The entrainment or mixing of material between plumes and inter-plumes is hindered by the
magnetic field. The smaller the magnetic energy density relative to the thermal or the kinetic
energy density,  the more efficiently entrainment takes place. Now consider the following numbers:
from $r=R_\odot$ to  $r=2R_\odot$ the density $\rho$ decreases by a factor of approximately $10^3$,
while the area of a plume  increases by roughly $\sqrt{10^3}$. Consequently the kinetic energy
density, $\rho v^2/2$, remains  roughly constant over this height range, while the magnetic energy
density, $B^2/2\mu$ drops by roughly a factor of $10^3$. This may explain why the plum properties
remain distinct from those of the inter-plume gas up to about $r=2R_\odot$ and then start to merge.
As far as the actual interaction is concerned, it may be enhanced by, e.g., the Kelvin-Helmholtz
instability which can act at the boundary between two  streams with sufficiently different speeds.
In addition electron collisions may not be sufficient at heights $\ge2.0~R_{\sun}$ to keep the
different species in equilibrium. Consequently, plume species are heated and accelerated
differently and rapidly as the density drops sharply with height. This may not be the case below
this altitude where the density is high enough to keep all species in equilibrium contrasting
inter-plumes regions where the (electron) density is smaller allowing for the plasma equilibrium to
brake down at lower altitudes and thus the hotter and faster plasma compared to plume regions.

We find that observed profile shapes, line widths and total intensities of the \ion{O}{6} doublet
and of Ly$\alpha$ are better reproduced when the influence of appropriate plumes is included. In
particular, the intensity ratios of the \ion{O}{6} doublet below $1.5~R_{\sun}$ are smaller when the
plume contribution is taken into account. This is due to the density enhancement in the plume
region. This result is in disagreement with the interpretation preferred by Gabriel et al. (2003 \&
2005) that such a reduction in the ratio is due to the Doppler dimming effect, which would imply a
higher outflow speed in plumes than in inter-plume regions. In our analysis, such high outflow
speeds in the plumes produce synthetic profiles whose narrow component is Doppler shifted towards a
profile wing, in contrast to the observations.

It is noteworthy that including polar plumes increases the agreement of the computed profiles
with the observed ones and with critical parameters like line width or \ion{O}{6} line ratio even if
no anisotropy in the temperature (or turbulence velocity) is included in the computations. Note that
the properties of the plumes were constrained basically using the line profile shape only and the
quantitative comparison with the line parameters and ratio was only carried out later. This
strengthens the thesis put forward by Raouafi \& Solanki (2004 \& 2006) that there is no compelling
need for large anisotropies in temperature in coronal holes based on published UVCS data.

Note that we have not tried to optimally reproduce the observed quantities by carrying out an
exhaustive search in parameter space. Thus the obtained results can possibly be improved further by
applying a $\chi^2$ minimization procedure to fit the line profiles.

Plumes with footpoints within $10\degr-15\degr$ of the pole display significant narrow components
even at altitudes above $\approx2.5~R_{\sun}$, contrary to the observations. EIT images during the
solar minimum together with high sensitivity magnetograms from SOLIS show evidence supporting the
hypothesis that plumes originate away from the pole. This subject will be studied in detail in a
subsequent paper.

\acknowledgments

The authors would like to thank Craig DeForest for constructive comments and critics that greatly
improved the manuscript and John Raymond for helpful discussions. The NSO is operated by the
Association of Universities for Research in Astronomy, Inc. under cooperative agreement with the
National Science Foundation. NER work is supported by NSO and the NASA grant NNH05AA12I.

\clearpage



\end{document}